# Optimizing Returns Using the Hurst Exponent and Q Learning on Momentum and Mean Reversion Strategies

Yejun (Andrew) Chang, Christian Lizardi, Rohan Shah

Cornell Data Science

May 21, 2022

**Abstract**

Momentum and mean reversion trading strategies have opposite characteristics. The former is generally better with trending assets, and the latter is generally better with mean reverting assets. Using the Hurst exponent, which classifies time series as trending or mean reverting, we attempt to trade with each strategy when it is advantageous to generate higher returns on average. We ultimately find that trading with the Hurst exponent can achieve higher returns, but it also comes at a higher risk. Finally, we consider limitations of our study and propose a method using Q-learning to improve our strategy and implementation of individual algorithms.

## 1. Introduction

Momentum and mean reversion trading strategies are two of the most commonly used algorithmic trading strategies. Momentum strategies identify recent trends in prices; traders generally buy when recent trends point upwards and sell when recent trends point downwards. Mean reversion strategies assess whether prices tend to revert to the mean over time; traders generally buy when prices fall below the mean and sell when prices rise above the mean. These two strategies are popular due to their relative simplicity and their ability to sustain net positive returns when implemented and optimized well [1].

There is an interesting contrast between these two widely used strategies: they display opposite characteristics [2]. Momentum strategies tend to lose frequently in small amounts with occasional huge payouts during price spikes, and they tend to generate better returns on assets that display persisting trends (i.e. assets that continuously and steadily grow). In contrast, mean reversion strategies tend to win frequently in small amounts with occasional huge losses when prices break out to the new mean, and they tend to generate better returns on assets that are mean reverting. To verify that these tendencies are well founded, we conducted thousands of trades with our implementation of momentum and mean reverting strategies. The distribution of returns that we obtained for each strategy is depicted below in **Figure 1**, which supports the characteristics discussed above. The distribution at left showing return rates for momentum strategy has a high frequency of negative returns, but positive returns can go up to 0.25. The distribution at right showing return rates for mean reversion strategy has a high frequency of positive returns, but negative returns can go as low as −0.20.

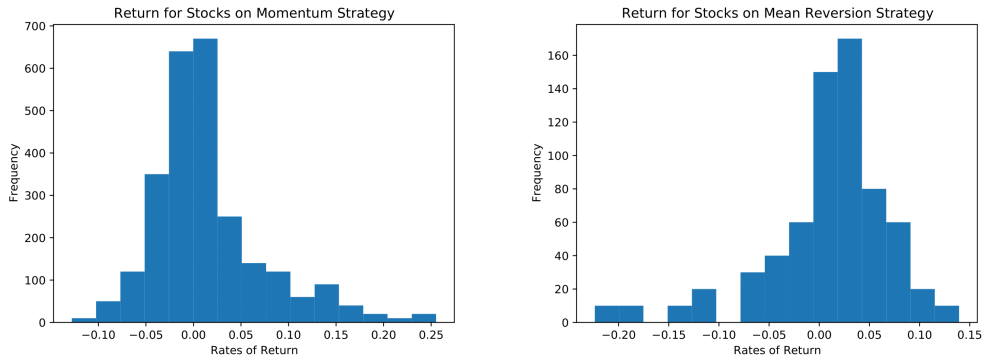

**Figure 1.** Return Rates on Momentum and Mean Reversion Strategies

The motivation behind this research is to take advantage of these opposite characteristics by trading with momentum on assets that are trending overall and trading with mean reversion on assets that are mean reverting overall. We achieve this by using the Hurst exponent, which classifies a time series data as trending, mean reverting, or a random series [3]. The domain of the Hurst exponent is between 0 and 1; time series is trending when the value is greater than 0.5 and mean reverting when the value is less than 0.5. Time series is a random walk when the Hurst exponent is equal to 0.5, but this case is not considered because we run a binary classification and the calculated value will never be precisely 0.5. With this understanding of the Hurst exponent, we trade with momentum strategies when the exponent is greater than 0.5 and with mean reverting strategies when the exponent is less than 0.5. Ideally, we anticipate this balanced strategy to increase expected returns by avoiding trading with momentum strategy on mean reverting assets and with mean reversion strategy on trending assets, which are considered disadvantageous due to characteristics described in the previous paragraph.

## 2. Strategy Implementations

### 2.1. Momentum: Moving Average Crossovers

We have implemented a simple moving average (SMA) crossover algorithm as our momentum trading strategy. The basis of SMA crossovers is to take shorter and longer term averages of recent prices and compare them. Let the shorter term average measure the mean price over the past five days, and let the longer term average measure the mean price over the past ten days, which are the parameters we chose for this research. If the shorter term moving average line is crossing the longer term moving average line from below, also known as the golden cross, it indicates an upward momentum and a buy signal. If the shorter term moving average line is crossing the longer term moving average line from above, also known as the death cross, it indicates a downward momentum and a sell signal likewise [4].

**2.2. Mean Reversion: Bollinger Bands**

A useful indicator for executing mean reverting strategies are the bollinger bands. Bollinger bands track moving averages, typically over the past twenty days, and the standard deviation of prices over the same period. Two curves are drawn two standard deviations above and below the moving average to form a band that indicates the likely range of price movements. Since bollinger bands are drawn at two standard deviations, 95% of price movements will be expected to fall within the band. When the asset price breaks out of the upper band, it indicates that assets have been overbought and are likely due to fall back to the mean. When the asset price breaks below the lower band, it indicates that assets have been oversold and are likely due to rise back to the mean again [5]. Traders may buy assets when the price breaks below the lower band, or wait until the price recovers back into the band for mitigated risk. Exit points can be targeted when the price crosses the moving average line or the upper band, which is up to the trader's discretion based on how much risk and reward they are willing to take [6].

**3. Preliminary Trial and Analysis**

**3.1. Methods**

1. From Nasdaq's stock screener, we compiled a list of stocks with market capitalization level at medium or above, which is defined as larger than $2 billion. The market cap restriction was put in place to filter penny stocks and other volatile stocks that are common at lower market cap stocks. Also, stocks with larger market cap tend to have been around for longer. Since the timeframe we analyze is 2021, filtering with market cap allows us to avoid having an abundance of incomplete data.
2. Download the daily time series closing price data for each stock, which ended up to be nearly 1800 Nasdaq stocks after filtering those with incomplete price data for 2021.
3. Find the Hurst exponent of each stock for the first half of 2021.
4. Trade with momentum strategy for all stocks for the second half of 2021.
5. Trade with mean reversion strategy for all stocks for the second half of 2021.
6. For Steps 4 and 5, find mean, median, and standard deviation of final returns.
7. For stocks with the Hurst exponent less than 0.5, take the final return from trading with mean reversion strategy. Likewise, for stocks with the Hurst exponent greater than 0.5, take the final return from trading with momentum strategy. Find mean, median, and standard deviation of selected final returns.

## 3.2. Results

|  | Momentum Returns | Mean Reversion Returns |
|---|---|---|
| Mean | 0.05% | 1.92% |
| Median | −0.16% | 1.64% |
| Standard Deviation | 2.90% | 6.48% |

**Figure 2.** Mean, median, and standard deviation obtained from Step 6.

**Figure 2** indicates that momentum strategy has lower mean, median, and standard deviation of returns compared to mean reversion strategy. This aligns with the tendencies addressed in the introduction: the lower and negative median indicates that momentum trading makes frequent losses in small amounts, and mean reversion trading makes frequent wins in small amounts in contrast. Lower standard deviation for mean reversion strategy implies that it is less volatile than momentum strategy. Generally, momentum strategy is considered riskier because the strategy inherently profits from volatility and trends are less self-sustaining than mean reversion. However, our analysis has been conducted over a 6-months horizon, which is relatively long term. With more time, positive trends may persist, and there is more opportunity for mean reverting patterns to break out to the new mean, which is a point where mean reverting strategies generally suffer huge losses.

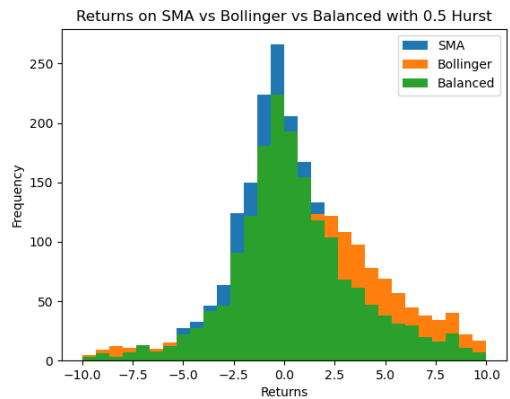

**Figure 3.** Histogram for returns. Blue distribution is when all stocks trade on momentum. Orange distribution is when all stocks trade on mean reversion. Green distribution is when the Hurst exponent method was applied.

**Figure 3** depicts a distribution of returns when the classification and selection using the Hurst exponent from Step 7 was applied. Contrary to the expectations, the mean was not higher than both blue and orange distributions. Instead, it lies in between the two distributions, and so does the standard deviation. The Hurst exponent strategy places expected returns and risk somewhere in between that of momentum strategy and mean reversion strategy.

## 4. Modification

### 4.1. Methods

**Figure 3** appeared as if the green distribution was in the middle of the blue distribution shifting down to the orange distribution. This observation prompted us to alter the boundary for the Hurst exponent where we divided which stocks traded on momentum and mean reversion strategies. Given Hurst exponent $x$ within the interval [0, 1], stocks with Hurst exponent less than $x$ traded on mean reversion, and those with Hurst exponent greater than $x$ traded on momentum instead of the default 0.5. This method can be somewhat counterintuitive as it goes against the Hurst exponent's definition. However, the Hurst exponent acts as a suggestion not a rule; just because a stock's Hurst exponent is less than 0.5 does not necessitate us to trade with mean reversion. The intent is to observe how mean, median, and standard deviation of returns vary as the boundary for the Hurst exponent increases from 0 to 1 and conclude if there is an ideal point of division that maximizes returns and minimizes risk.

### 4.2. Results

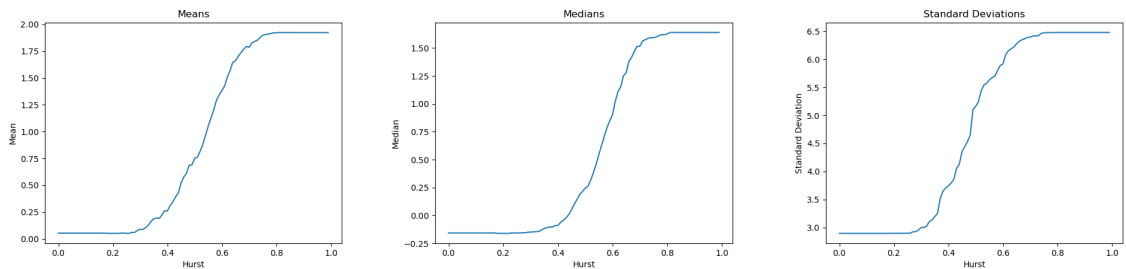

**Figure 4.** Means, medians, and standard deviations of returns vs Hurst exponent value that divides whether a stock trades on momentum or mean reversion

All three curves in **Figure 4** display a logistic trend. As the Hurst exponent boundary increased from 0 to 1, all mean, median, and standard deviation of returns increased. When trading more stocks with momentum (i.e. the Hurst exponent boundary is less than 0.5), one would expect lower returns on average but also lower risk. When trading more stocks with mean reversion (i.e. the Hurst exponent boundary is greater than 0.5, one would expect higher returns on average but also higher risk.

## 5. Conclusion

From the preliminary trials, we found that using the Hurst exponent to make trading decisions does not necessarily increase returns. The modified trial sliding the Hurst exponent revealed that the Hurst exponent strategy does not only influence returns but also risk, which adds complexity to the strategy. We conclude that using this Hurst exponent strategy is not mainly about improving returns. Instead, it is more about choosing the level of risk and reward that one is comfortable with.

### 5.1. Limitations

The main goal of this research was not to optimize each momentum and mean reversion strategy, but to balance the two strategies and optimize the total returns. Hence, we implemented simple yet serviceable strategies, SMA crossover and bollinger bands, as our choice for momentum and mean reversion trading algorithms. However, going back to **Figure 1**, the expected returns for both algorithms are quite low. Assuming the risk free rate of 2%, the Sharpe ratio would be negative in both cases. The final trends and outcomes we have obtained are somewhat restricted to the versions of algorithms we have implemented and cannot necessarily be generalized to different and relatively optimized versions of momentum and mean reverting strategies. Furthermore, stocks are not strictly trending or mean reverting; for instance, mean reversion can be contained within long-term trends with moving averages, so binary classification of time series data into trending or mean reverting might not be the best idea.

## 6. Further Steps: Optimizing Returns Using Q Learning

As addressed in the limitations section above, the results from our trials call for improvements to our strategy implementations. We ran a separate experiment to try optimizing our strategies with Q-learning. The goal is to be able to obtain more favorable return distributions that we could potentially balance in a similar way to how we balanced the simpler trading algorithms above.

Q-learning is a type of reinforcement learning algorithm that develops Q estimates which encode the expected value of taking particular actions in particular states. The algorithm learns the expected payoffs of various action state pairs over a training period. At the point of the reward estimation function's convergence or at the end of the training data, we aim to have created Q estimates that can inform a fully capable actor in the environment that we capture using a state space.

In the case of a trading algorithm, we intend to create a fully capable trader that is able to learn the trading environment using its representation of the world in terms of a state space, actions, and rewards. Because the underlying model and hyperparameter optimization of the model will be similar regardless of the type of strategy we investigate, we only focus on momentum trading algorithms. The software and analysis we develop for momentum could easily be adjusted to other types of strategies by adjusting the state space, which is the information that the algorithm considers when making an action. Furthermore, we focus on a particular state that attempts to inform the algorithm of trends; however, there are many other potential trend-capturing states that could be used as the baseline for a momentum trading algorithm.

We used a state space based on the relative value of the short-term moving average minus the long term moving average at time $t$ and that same value at time $t-1$. We discretized the state space to group similar states and be able to customly adjust the cardinality of the state space. The defined state space allows the reinforcement learning algorithm to use similar reasoning as the simple moving average cross algorithm that we have implemented for this study. The results were promising, as when the algorithm restricted its actions so that it only acted when it expected a reward with statistical significance, it could achieve high returns and sharpe ratio on many equities. When it was forced to act in every state, it performed poorly. This implies that not every state gives a signal that is worth trading on. Note that the risky actions we allowed it to consider taking on each day were to hold or short from the beginning to the end of the day. It will be worth exploring other kinds of actions in the future, such as holding positions for longer periods. Lastly, when training the algorithm on larger collections of equities, it was able to achieve higher sharpe ratios and the results were more consistently strong. This potentially implies that the algorithm was able to find more consistent strong strategies when trained across multiple equities or over more training data.

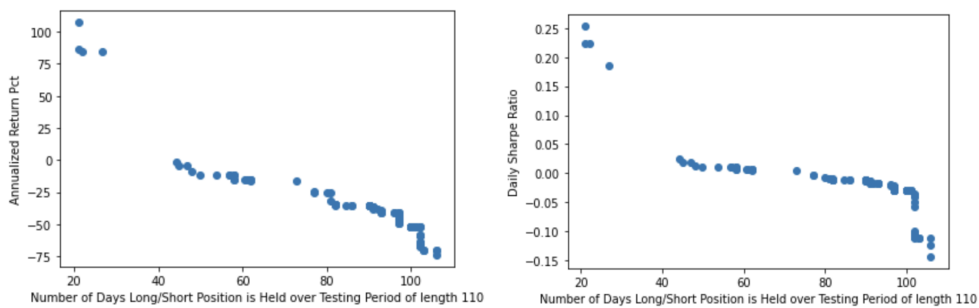

**Figure 5.** Training on one equity had more volatile returns leading to lower sharpe ratios.

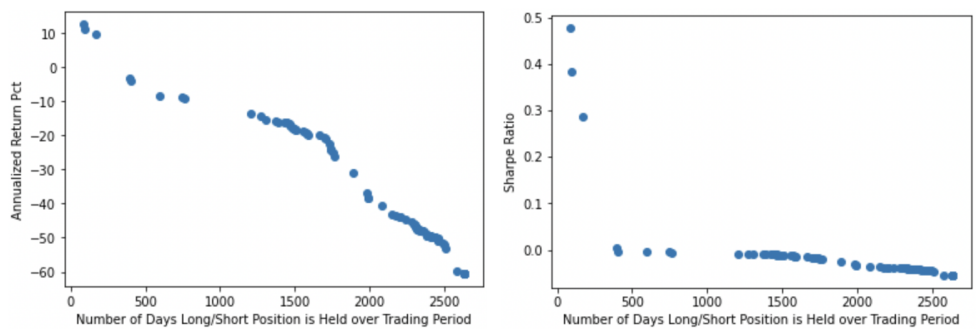

**Figure 6.** Training on many equities generally had more consistent returns leading to higher sharpe ratios.

We will further investigate this algorithm in the future. We hope to obtain more definite cause and effect relationships as well as determine to consistently train algorithms that are capable of achieving higher sharpe ratios using this state space. Additional steps include investigating other momentum-based state spaces as well as state spaces based on different types of strategies, and extending the application of this Q-learning strategy to mean reversion to optimize both strategies. Lastly, once we obtain collections of stronger momentum and mean reverting algorithms, it would be interesting to balance their distributions using the Hurst exponent.

## Citations